\documentclass[aps,article,reprint,groupedaddress,superscriptaddress,amsmath,amssymb]{revtex4-1}[10pts]

\usepackage{amsmath}
\usepackage{amssymb}
\usepackage{multirow}
\usepackage{graphicx}  
\usepackage{dcolumn}   
\usepackage{bm}        
\usepackage{verbatim}   
\usepackage{lipsum}
\usepackage{graphicx}
\usepackage{braket}  
\usepackage{ulem}
\usepackage{soul}
\usepackage{color}
\usepackage{float}
\restylefloat{table}
\usepackage{physics}
\newcommand{\tb}{\textcolor{magenta}}

\begin{document}

\title{Quantum stabilization of microcavity excitation in a coupled 
microcavity--half-cavity system}
\author{C.~Y.~Chang}
\email{cychang@gatech.edu Present address: cychang@meso.t.u-tokyo.ac.jp} 
\affiliation{Georgia Institute of Technology, School of Physics, Atlanta, Georgia 30332-0250 USA}%

\author{Lo\"{i}c Lanco}
\affiliation{Centre de Nanosciences et de Nanotechnologies, CNRS, Université Paris-Sud, Université Paris-Saclay, C2N, Marcoussis, 91460 Marcoussis, France}
\affiliation{Université Paris Diderot, Paris 7, 75205 Paris CEDEX 13, France}


\author{D. S. Citrin}
\affiliation{School of Electrical and Computer Engineering, Georgia Institute of Technology, Atlanta, Georgia 30332-0250 USA}%

\date{\today}

\begin{abstract}
We analyze the quantum dynamics of a two-level emitter in a resonant microcavity 
with optical feedback provided by a distant mirror 
(\textit{i.e.}, a half-cavity) with a focus on stabilizing the 
emitter-microcavity subsystem. Our treatment is fully carried out in the framework of 
cavity quantum electrodynamics. Specifically, we focus on the dynamics of a perturbed 
dark state of the emitter-microcavity subsystem to ascertain its stability (existence of time 
oscillatory solutions around the candidate state) or lack thereof.  
In particular, we find conditions under which multiple feedback modes of the half cavity 
contribute to the stability, showing certain analogies with
the Lang-Kobayashi equations, which describe a laser diode subject to classical optical 
feedback.

\end{abstract}

\maketitle

\section{introduction}

With recent advances in the fabrication of nanophotonic structures, 
there is an increasing ability to control and manipulate the optical properties 
on the single-photon 
level \cite{Kimble2003ExperOneAtomLaser,lodahl2015interfacingRMP, reiserer2015cavity}. 
One of the techniques central to such control is the ability to access the strong-coupling 
regime of cavity quantum electrodynamics (cQED). In solid-state structures, 
recent developments include coupling quantum emitters such as quantum dots (QD)
to photonic crystals  \cite{lodahl2004controlling,Lodahl2015Ncomm,nomura2009laser} 
and micropillar cavities \cite{bockler2008electrically, gazzano2013bright}. 
These devices provide unidirectional photons that are of interest, for example, to improve 
quantum communications over long distances \cite{bouwmeester1997experimental}. 
In addition, the ability to control and coherently manipulate photons coupled to the internal state 
of the QD is of great importance due to potential  applications in quantum memory and
 quantum information processing \cite{giesz2016coherent}.

Controlling a quantum system--in the sense of providing stabilization of its
quantum state--presents a nontrivial problem. The use of feedback to provide 
stabilization in classical systems is quite advanced, while less is known for quantum system. 
Two feedback-control schemes have been 
explored recently in quantum optics, \textit{viz.}\  measurement-based and coherent feedback 
loops \cite{serafini2012feedback, wiseman2009quantum}. 
In a measurement-based feedback loop, the quantum system is monitored and the 
outcome of the measurement is used as \textit{classical} information 
to manipulate the operations. 
A measurement-based feedback loop has been implemented in a 
cQED system with 
trapped atoms \cite{WisemanPhysRevLett.70.548}.  
In coherent feedback, the dynamics are entirely quantum mechanical and
the system interacts coherently with an ancillary subsystem 
in both the extraction and manipulation processes. 
For coherent feedback, such as Pyragas feedback \cite{PYRAGAS}, 
both processes utilize information 
stored in the reservoir degrees of freedom residing, for example, in an external cavity (EC).

\begin{figure}[b]
    \centering
    \includegraphics[width=0.48\textwidth]{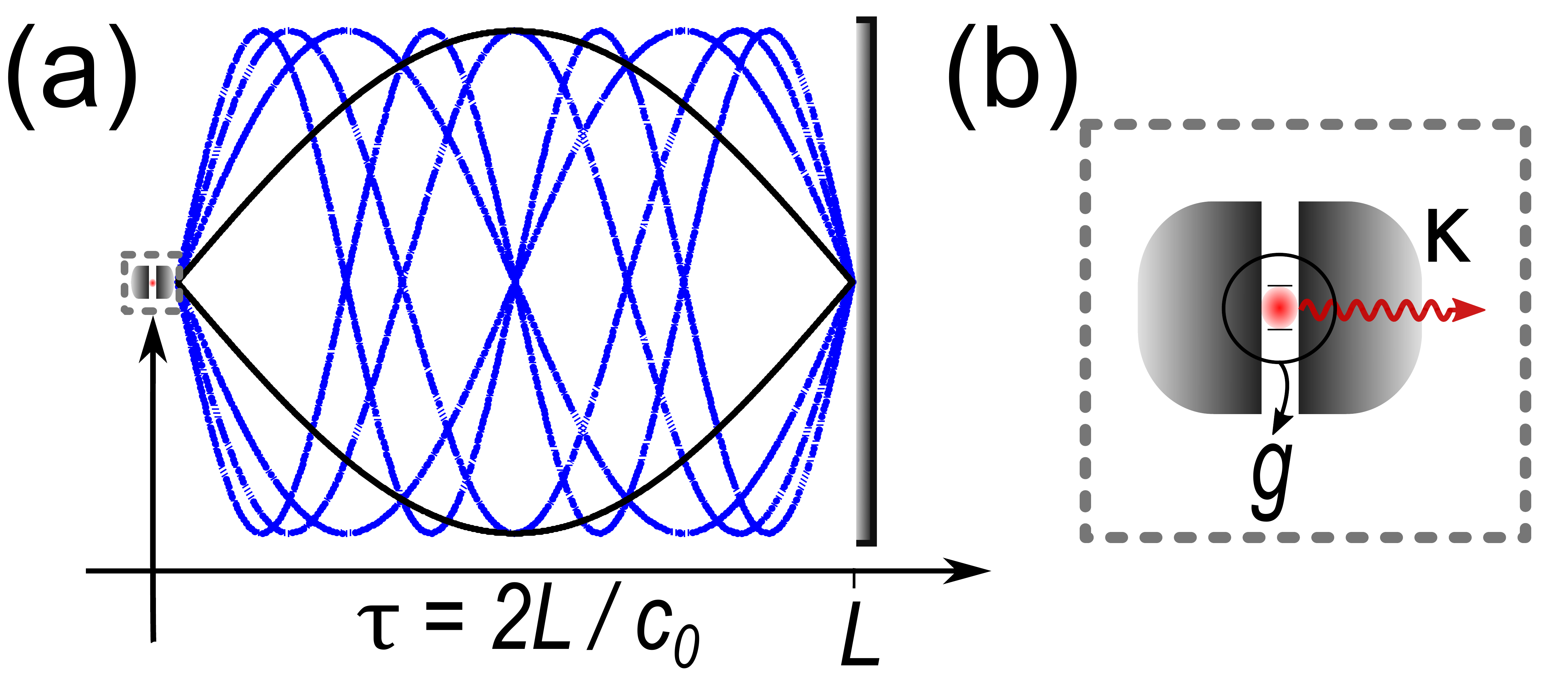}  
    \caption{(a) The \tb{quantum-dot-micro-cavity} system coupled to  an  EC of length 
    $L$ in which a quasi-continuum of photon modes exist.     
    $\tau$ is the delay time of the coherent feedback. The black and the blue curves indicates classical electric fields forming standing waves from the external cavity. (b) Schematic of a two-level QD in a near-resonant MC, 
    where $g$ is the coupling strength and 
    $\kappa=\pi G_0^2/(2c_0)$
    is the MC damping rate. 
    }
   \vspace{-15pt}
   \label{system}
\end{figure}

\subsection*{Canonical consideration }
Given that the measurement process will unavoidably collapse the wavefunction, 
measurement-based feedback presents limitations when the target state is a superposition. 
To implement coherent quantum feedback in quantum optics, 
one typical configuration consists in coupling the quantum system with an external 
half-cavity, having a perfect mirror at the opposite end, as 
illustrated in Fig.\ \ref{system}(a), similar to the time-delayed 
feedback setup used in EC semiconductor 
lasers (ECSL). 
The dynamics of ECSLs have been intensively studied based on the 
semiclassical
Lang-Kobayashi (LK) 
equations \cite{lk1070479}. The LK equations describe the nonlinear dynamics 
of the electric-field amplitude $|E|$, carrier density $n$ in the active region, and optical phase 
$\Delta\phi$ by a set of equations of motions (EOM) in the form of coupled delayed-differential 
equations. The dynamics are described by trajectories in phase space spanned by the 
three dynamical variables.

The presence of time-delayed feedback is known for the ECSL to generate 
multiple steady-state solutions. The stable modes of the EC are called EC modes
(ECM), while 
the unstable modes of the EC are known as  
antimodes \cite{EllipseLKJQE}. Depending on the various 
parameters of the system (e.g., 
injection current, feedback strength, optical feedback phase $\Delta\phi$, 
and EC length $L$), various dynamics can occur 
near and amongst these solutions \cite{wolfrum2006eckhaus,green2010bifurcation}. 
In particular, oscillatory dynamics about various ECMs occur in certain parameter regimes,
and are closely related to undamped relaxation oscillations
\cite{erneux2007time}, while more complex behavior exhibiting 
closed trajectories encircling  one or more steady-state solutions are also observed. 
The emergence of multiple solutions is 
often seen in many nonlinear systems, but not in a quantum system with a finite 
number of degrees of freedom \cite{eckhaus1965studies}. 
However, in quantum systems
with a bath of infinite degrees of freedom (in our case, the modes of the EC), 
once these are integrated out, the system may exhibit
what appears to be nonlinear behavior in the remaining explicitly considered degrees of
freedom.

Another point of view of nonlinearity in quantum systems with coherent 
feedback is to consider
the variation in the dynamical behavior with the number of photons. 
The nonlinearity involving states 
with a small number of photons has been observed experimentally \cite{hopfmann2013NJPnonlinear,QDchaosalbert2011}. 
These effects cannot be described with a classical electromagnetic field, \textit{i.e.}, 
the scaling of the behavior with photon number cannot be accounted for by the 
corresponding optical intensity. Hence, a fully quantum mechanical treatment of such 
time-delayed systems, including the electromagnetic field, is essential to understand the 
dynamics of single-photon emitters with coherent time-delayed feedback.

Broadly speaking, two approaches have been used to account for the coherent feedback. 
In one approach, the non-Markovian evolution of the coherent feedback effect is analyzed with a differential equation with a time-propagator \cite{grimsmo2015time}. On the other hand, the optical feedback from a  mirror has typically been accounted for using a one-dimensional model for the electromagnetic field with 
a standing-wave basis in the Markovian limit \cite{CookPhysRevA.35.5081,ZollerPhysRevA.66.023816,mirrorPhysRevA.66.063801}. 
Recent theoretical studies have focused on Markovian coherent feedback in various systems, 
such as ensembles of 
atoms \cite{CookPhysRevA.35.5081,ZollerPhysRevA.66.023816,mirrorPhysRevA.66.063801}, 
a single atom in a cavity \cite{DChang2012cavity}, one-dimension waveguides 
\cite{ fang2015waveguide, guimond2017delayed, IHoiNature, TommasoPhysRevA.90.012113, calajo2018}, and photonic 
crystals \cite{PhysRevLett.110.013601, kabuss2015analytical, zoller2016photonic}. 
It has been demonstrated that the standing-wave field model in a half cavity 
(Markovian) is equivalent to a time-delayed feedback system 
(non-Markovian) \cite{mirrorPhysRevA.66.063801,kabuss2015analytical}. 
The control of the delay time can be a key  factor in the stability or instability as
well as the nonlinear dynamics in many complex 
systems \cite{RevModPhys.85.421,grimsmo2014rapid}, providing wide-range applications. 
These works provide a promising route to rapid convergence of 
states \cite{grimsmo2014rapid}, enhancing entangled photon-pair generation from 
biexcitons \cite{BiexcitonPhysRevLett.113.027401}, and to drive continuous 
exchanges for pure states \cite{PhysRevLett.110.013601}. Among these studies, 
stability in coherent feedback systems has been investigated by 
Grimsmo \cite{grimsmo2015time} based on linear delayed differential equations (non-Markovian). 
Since then, few other studies have focused on stability and its relation to 
time-delayed systems \cite{NemetPhysRevA.94.023809,tabak2016trapped}.
The evolution of the quantum states in these investigations is described by means 
of a time-varying Hamiltonian. These works are focused on specific features of the 
quantum system and not on the stability of the target state, which from the 
standpoint of experimental realization or practical application is of key importance.
 
In this study, we consider a quantum system 
composed of a single-photon emitter in the form of a quantum dot (QD) within a 
microcavity (MC) formed by a micropillar with coherent quantum feedback provided by a 
distant mirror as shown in Fig. \ref{system}(a). Specifically, in view of the relative lack of understanding of how quantum feedback can provide stability in quantum systems, we focus on analyzing the stability of such system in various parameter spaces. Thus, this work provides a  testbed to study time-delayed 
feedback in a fully quantum system, as well as addresses a physical
implementation of such a system of interest for nanophotonic and quantum
information-science applications. Our approach presents a direct stability analysis 
in the Markovian dynamics of quantum feedback from a single-photon emitter.
The quantum feedback is achieved via the EC, 
similar to previous studies \cite{PhysRevLett.110.013601, kabuss2015analytical}. 
We derive a set of EOMs for the quantum amplitudes associated with the 
natural basis describing the QD excitation, the MC photon, and the EC modes. 
The EOMs of the state vector are obtained with the input-output formalism 
in the Schr{\"{o}}dinger picture instead of in the Heisenberg picture as in 
Ref.\ \cite{zoller2016photonic}.

Since photon leakage from the EC is neglected, our system conserves excitation number. 
Our focus is on the one-excitation subspace.  In the one-excitation subspace, 
we find the stationary-state solutions, which we then investigate for stability. 
By \textit{stationary state},
we mean states in which the probabilities in the noninteracting basis 
remain constant in time even including the interaction. 
In other words, stationary states are those that are simultaneously 
eigenstates of both the noninteracting and the interacting Hamiltonians. 
We explore the stabilty of the stationary states, that is, states of the composite
MC/QD subsystem that are effectively decoupled from the EC.
We find that the \textit{stable} stationary states are represented by the 
MC/QD and its mirrored-self being in a singlet state \cite{vetter2016single}. 
Similar observations have been made that the singlet state 
corrresponds to Dicke subradiant state \cite{DickePhysRev.93.99} in 
 a waveguide system  \cite{DChang2012cavity}, in a system 
with chiral feedback from a single V-level atom \cite{ZollerChiralPhysRevA.94.033829}. To be more specific, 
the singlet state would appear effectively 
decoupled from the waveguide, being in a dark state which is subradiant, 
and thus stable, whereas the triplet state is superradiant, and thus unstable, 
as it decays at 
twice the cavity damping rate from the 
MC, \cite{DChang2012cavity,ZollerChiralPhysRevA.94.033829}. 
In the case with one single-photon emitter coupled to the EC, the subradiant (dark) state would be the only stationary state where the population in QD and MC are antisymmetric \cite{DickePhysRev.93.99}. 

We determine the stability by studying the dynamics in the vicinity of the 
stationary state. This is done by constructing the Jacobian matrix of the linearized 
EOMs with the state amplitudes perturbed from the stationary 
state \cite{eckhaus1965studies}, a technique widely employed for classical stability analysis.
We analyze the eigenvalues of the Jacobian matrix as a function of 
coupling strength between the MC and EC. 
A strictly imaginary eigenvalue indicates oscillatory dynamics about the 
candidate stationary state, which is
thus deemed stable while a positive (negative) \textit{real part} of the 
complex eigenvalues indicates unstable (stable) stationary states.
The effects of time delay are also studied for various EC length $L$. 
We numerically verify the Jacobian 
analysis by perturbing the stationary state to investigate its stability, 
which agrees with the results obtained directly from the Jacobian of the state amplitudes. 

The remainder of the paper is organized as follows.  In the next section, we outline the cQED description of the system.  We next find the stationary state.  
Following this, we assess the stationary state's stability.  In the final section, we conclude.
\\
\section{Theoretical description of the single-emitter, microcavity, external cavity system}

We begin by introducing the system.  We consider an intrinsic QD coupled to a single 
near-resonant
mode of a high-$Q$ micropillar MC [Fig.\ \ref{system} (b)] and coupled in turn
to the EC modes shown in Fig.\ \ref{system}(a). The QD is characterized by 
interband-transition frequency, $\omega_0$. The QD interband transition is 
dipole coupled to a single mode 
of the micropillar MC 
of angular frequency $\omega_{MC}$ (in this study, we will eventually 
take $\omega_{MC}=\omega_0$)
with  coupling strength $g$, 
as in Ref. \cite{somaschi2016near}. This approach can yield strong coupling 
between the QD and the MC \cite{BiexcitonPhysRevLett.113.027401} and 
can generate high-purity, indistinguishable single photons \cite{somaschi2016near}. 
We place an ideal mirror with reflection coefficient $r\!= \! -1$ a distance 
$L\!=\! c_0\tau/2$ 
from the micropillar to form the EC (half cavity), with $c_0$  
the speed of light in vacuum and $\tau$ the EC round-trip feedback time. 
The conditions we choose are similar to the single QD in a MC subjected 
to an external mirror in a recent paper \cite{kabuss2015analytical}. 
For a mirror with $|r|\!<\!1$, the coherent 
feedback can be treated considering the mirror properties \cite{faulstich2018unraveling}, while analysis can be perform using an open quantum-system formalism, discussed in a recent publication by Whalem \cite{whalen2017open}.

We work in the rotating-wave approximation (RWA) \cite{walls2007QObook} 
in the Schr{\"{o}}dinger picture as in Refs.
\cite{CookPhysRevA.35.5081,PhysRevLett.110.013601,kabuss2015analytical}. 
The derivation of the interaction Hamiltonian can be found in Ref.
\cite{CookPhysRevA.35.5081,PhysRevLett.110.013601,kabuss2015analytical,quantumoptics,walls2007QObook,nicolasspie, zoller2016photonic} 
(see supplementary material). For the EC, we use the free-space 
dispersion for photons, $\omega_k \!=\! c_0 \abs{k}$, 
assuming a sufficiently large 
value of
$L$ so that the photon modes can be considered a quasi-continuum. 
We assume that only the optical modes with angular frequencies near $\omega_0$ 
interact strongly with the MC photon ($\omega_k\! \approx\! \omega_{MC}\!
\approx\!\omega_0$).
We obtain the following interaction Hamiltonian describing the QD coupled to the MC mode 
and that mode in turn coupled to the EC modes,
\[
\frac{H_{int}^{(RWA)}}{\hbar} \! = \! -  \omega_g (\sigma^-a^\dagger \! + \! \sigma^ + a)\! - \!\!
 \int_{-\infty}^\infty \!\!\!\! \!d\omega_k [G(k,t) a^\dagger b_k\!+\! \rm{h.c.}]   \label{1} \tag{1}
\]
with $\omega_g=g/\hbar$, $a^\dagger$ ($a$) the creation (annihilation) operator for the MC photon and 
$\sigma^{+(-)}$  the raising (lowering) operator of the 
two-level QD system. The bosonic operators $b_k$ 
destroy a photon of frequency $\omega_k$ in the EC as defined as 
in Ref.\ \cite{zoller2016photonic} and the coupling element, 
$G(k,t)= G_0\sin(kL)e^{i(\omega_0-\omega_k)t}$, where 
$G_0 = \sqrt{2c_0\kappa/\pi}$.
Note that because $H_{int}^{(RWA)}$ is quadratic in creation and annihilation operators, 
the Hamiltonian conserves the number of excitations. 
We extend the lower limit of integration to $-\infty$ given the interaction  
bandwidth is narrow compared with  $\omega_0$. A detailed discussion is presented 
in the supplemental material.

We now concentrate on the one-excitation subspace where an arbitrary wavefunction can be written
\[
\Psi(t) \!=\! c_{e}(t)\ket{e,0,0}\! +\! c_{c}(t)\ket{g,1,0}\! + \!
\int_{-\infty}^\infty \!\!\!\! c_{k}(t)\ket{g,0,k} dk.  \label{2} \tag{2}
\]
Here, in kets $\ket{a,b,k}$, $a\!=\!e$ ($g$) denotes the QD in the excited (ground) state; 
$b\!=\!0$ (1) denotes zero (one) MC photon; and $k$ denotes the 
EC photon wavevector. $\Psi(t)$ is thus described by the time-dependent amplitudes 
$c_e(t)$, $c_c(t)$, and $c_{k}(t)$.

\section{stationary state}

Substituting $\Psi(t)$ into the time-dependent Schr{\"{o}}dinger's equation, 
we obtain the following coupled EOMs for the time-dependent amplitudes,
\begin{align}
\frac{\partial c_{e}}{\partial t} =& \;  i \omega_g\; c_{c}, \label{3} \tag{3} \\
\frac{\partial c_{c}}{\partial t} =& \;  i \omega_g\; c_{e}  + 
i  \int_{-\infty}^{\infty} c_{k}~ G(k,t)\; \mathrm{d}k , \label{4} \tag{4} \\
\frac{\partial c_{k}}{\partial t} =& \; i G^{*}(k,t)\; c_{c}  . \label{5} \tag{5}
\end{align}

To find the stationary states (\textit{stationary} in the more narrow sense defined above), 
we write $c_{i}(t)=| c_i(t)| \exp [i\theta_i(t)]$, to obtain a new set of EOMs,

\begin{align}
 \frac{\partial c_{i}}{\partial t}=\frac{\partial \left| c_{i}\right|}{\partial t} e^{i\theta_i(t)}
 +i\frac{ \partial \theta_i(t)}{\partial t}  \left|c_{i}\right|e^{i\theta_i(t)}, \label{6} \tag{6}
\end{align}
with $i\! =\! e$, $c$, and $k$. We apply the constraint 
of stationarity $\partial\abs{c_i}/\partial t\! =\! 0$ and find 
the solution as discussed in the supplemental materials. 
 
  \begin{figure}[b]
  \vspace{-5pt}
    \centering
    \includegraphics[width=0.48\textwidth]{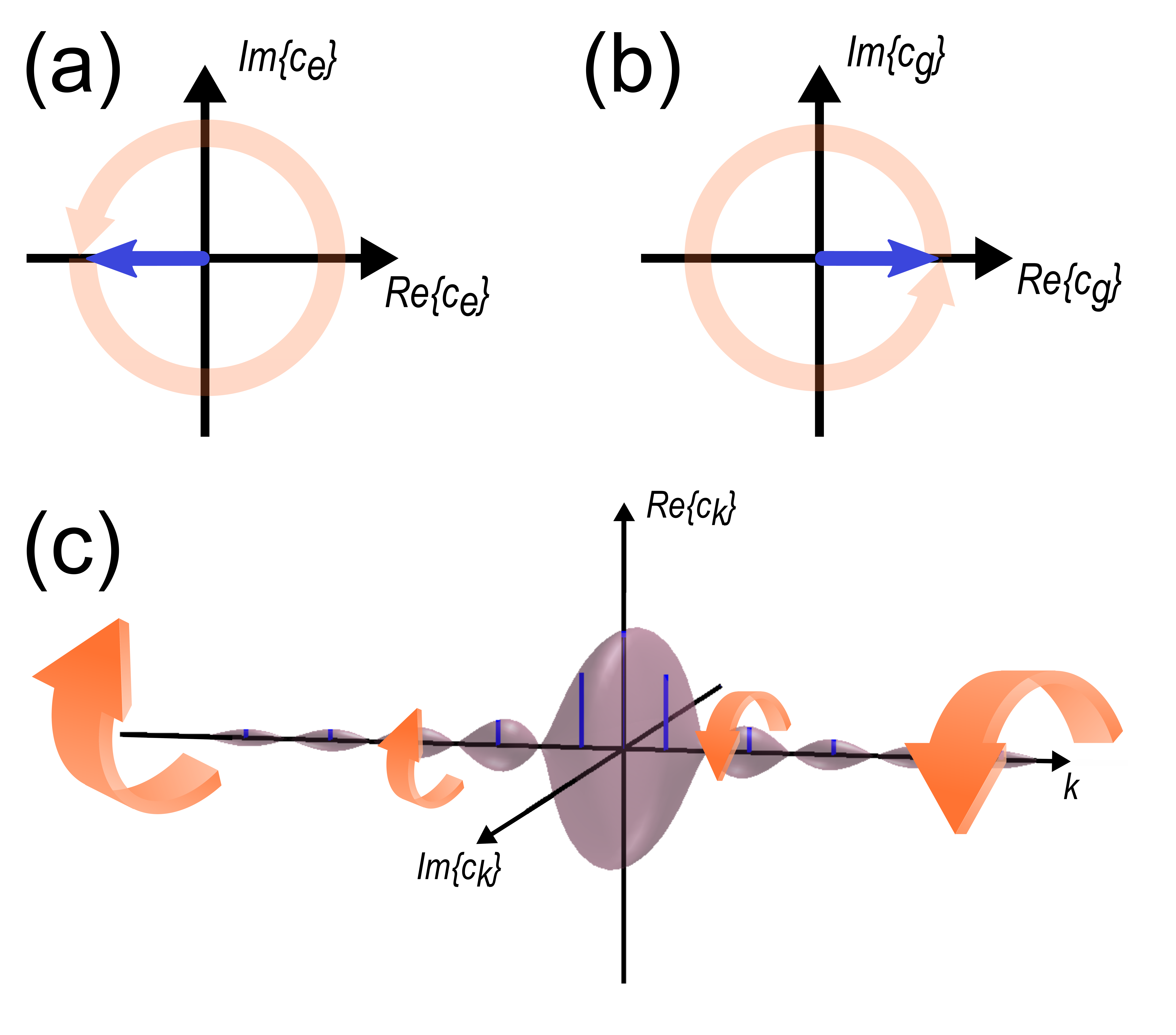}  
       \vspace{-10pt}
    \caption{Representation of the one-excitation stationary state 
   in the frame rotating at the frequency $\omega_0$ (a) $\bar{c}_e(t)$, (b) 
    $\bar{c}_c(t)$, 
    and (c) $\bar{c}_k(k,t)$. The blue arrows indicate the initial condition
    and the orange arrows indicates the time evolution.}   
   \vspace{-3pt}
   \label{state}
\end{figure}

We denote the stationary state (with respect to the interaction Hamiltonian) 
in the single-excitation subspace as $\bar{\Psi}(t)$ 
(where the bar indicates a stationary solution). In the frame rotating at $\omega_0$, $\bar{\Psi}_{RWA}(t) = \bar{\Psi}(t) e^{-i\omega_0t} $
\begin{flalign}
 \bar{\Psi}_{RWA}(t) =  \bar{c}_e(t)\ket{e,0,0} + & \; \bar{c}_c(t)\ket{g,1,0} \notag \\
 +&
 \int_{-\infty}^\infty  \bar{c}_k(t)\ket{g,0,k}dk   . \notag
\end{flalign}
\noindent 
We thus obtain the  noninteracting amplitudes for the stationary state
\begin{flalign}
\bar{c}_e(t) &= -\alpha e^{-i\omega_gt} , \label{7} \tag{7}\\
\bar{c}_c(t) &= \alpha e^{-i\omega_gt}, \label{8} \tag{8}\\
\bar{c}_k(k,t) &= \frac{\alpha G_0\sin(kL)}{\omega_k-\omega_g- \omega_0}
e^{i(\omega_k  - \omega_g - \omega_0)t}  . \label{9} \tag{9}
\end{flalign}

Note that the strict commensurability condition, \textit{i.e.,} $2L/c_0\! = n \;\cdot\; \!2\pi /\omega_g$ with $n\in \mathbb{N}$, is required for the stationary state. Detailed reasons will be explained later in this work 
and can also be found in Ref. \cite{fang2015waveguide, kabuss2015analytical}.
The parameter $\alpha$ is defined in the next paragraph.
Note that the minus sign results from the MC/QD being in a  singlet state (antisymmetric) of the quantum system and its mirrored self \cite{DChang2012cavity, vetter2016single,ZollerChiralPhysRevA.94.033829, mirrorPhysRevA.66.063801}. 
In this state, the QD-MC subsystem are effectively decoupled from the EC, 
thus being in a dark or subradiant state, as discussed for similar systems in Refs.\ \cite{DChang2012cavity,ZollerChiralPhysRevA.94.033829,vetter2016single}.

The noninteracting amplitudes associated with this stationary state can be 
characterized by a single parameter 
\[
 \alpha \!=\! \left|c_c(t)\right|\! = \!\left|c_e(t)\right|\! = \!
        \left(2+\frac{\pi  G_0^2L}{c_0^2}\right)^{\!\!-1/2} \!=\!(2+\tau\kappa)^{-1/2}
\]
where recall $G_0 = \sqrt{2c_0\kappa/\pi}$ and $\kappa$ is the MC photon 
damping rate. Since we consider 100 \% reflection 
from the distant mirror, the damping rate characterizes both the 
rate \textit{into the EC} and \textit{the feedback rate from the reflected photon.}

The stationary state is indicated in the frame rotating at $\omega_0$ in Fig.\ \ref{state}.  The blue arrow 
shows the state of the QD at a snapshot in time with the time evolution 
in the RWA shown by the orange arrows where we 
choose initial conditions 
$\bar{c}_e(0)\!=\!\alpha$, $\bar{c}_c(0)\!=\!-\alpha$, and 
$\bar{c}_{k}(k,0)\! =\! \alpha G_0\sin(kL)/(\omega_k-\omega_g)$.
In this case, the EC photon is in a standing wave 
and its population is described by a sinc function in the $k$ channel (of frequency $\omega_k$), centered on $\omega_k =\omega_g$, i.e. $k L = n \pi$, with $n$ an integer number such that the commensurability condition is fulfilled : commensurability condition $\tau = n \tau_g$ with $\tau_g =  2 \pi / \omega_g$, the time it takes for the strongly-coupled exciton-photon system to perform a full Rabi oscillation.   Note that each EC state, $c_k$, revolves around the $k$ axis at different rate, as illustrated in Fig. \ref{state}(c).

\begin{figure}[t]
    \centering
    \includegraphics[width=0.5\textwidth]{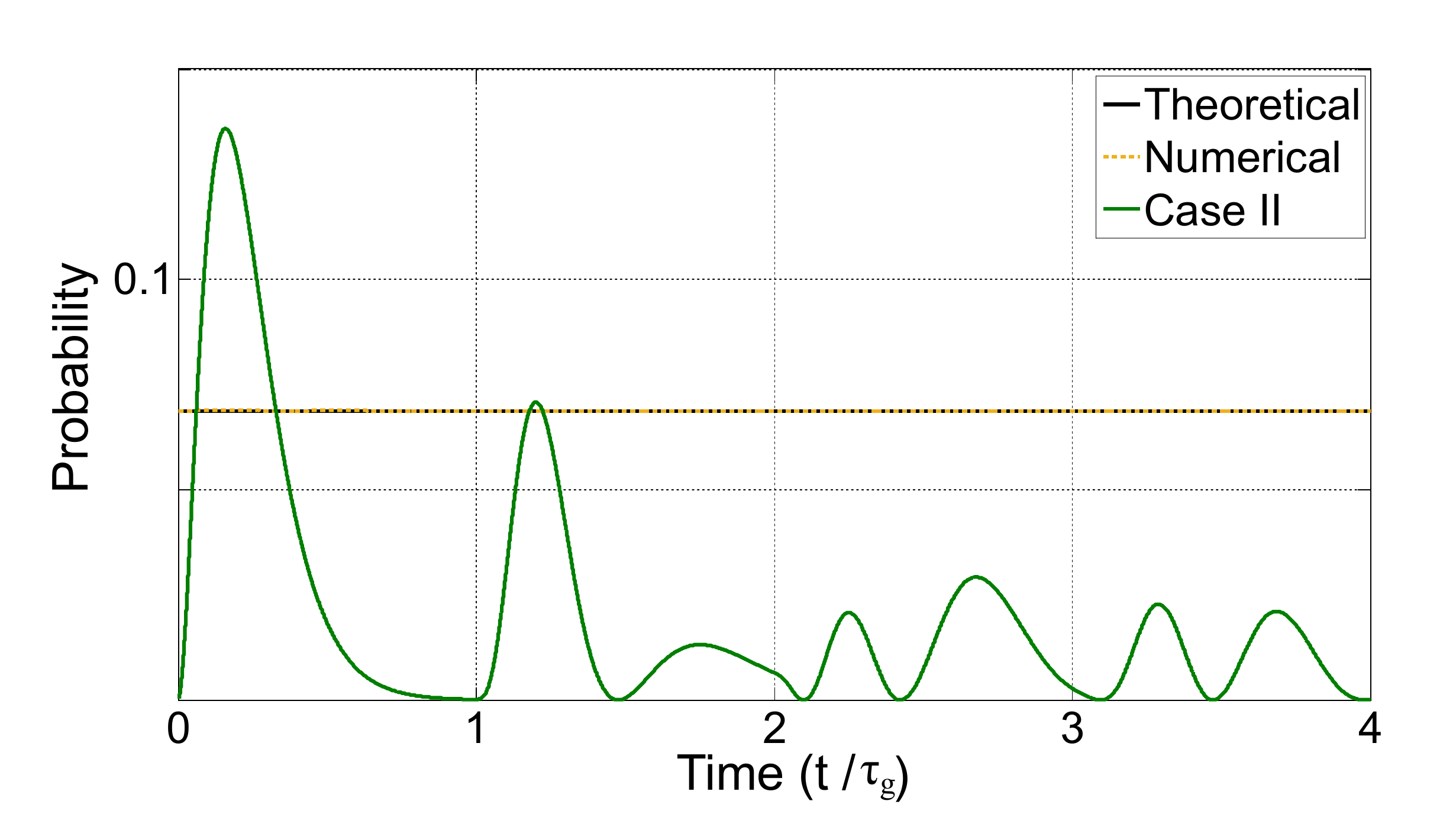} 
        \vspace{-10pt}
    \caption{Time evolution of the probabilities of $\abs{c_c(t)}^2$ for various initial conditions. 
    The QD is initialized in the excited state for the green curve
    (Case II); however,
      choosing  initial conditions corresponding to a stationary state 
    results in $\abs{c_c(t)}^2$ being independent of time (black and dotted yellow curves).}  
    \label{state_com}
       \vspace{-5pt}
\end{figure}

The evolution of the MC photon population $\abs{c_c(t)}^2$ is plotted in Fig. \ref{state_com}.
The theoretical and numerical results 
for the stationary state just found
are plotted in the black and yellow dotted curves, respectively.
In this case, the MC photon population remains constant in time.
The green curve (Case II)
plots the MC photon population $\abs{c_c}^2$, however, for the initial state 
$c_e(0)\!=\!1$, $c_c(0)\!=\!0$, $c_k(0)\!=\!0$ for all $k$.
In this case, the MC photon population varies significantly at fist and eventually mainly leaks 
out into the EC.  Note that there is only one stationary state in our sense within 
the one-excitation subspace.

To lend insight into the nature of the stationary state, 
the initial state amplitudes of 
the EC photons of wavenumber, $\bar{c}_k(k,0)$ are given by an \textit{even} function, centered on $\omega_k= \omega_g$, i.e. $k L = n \pi$  as in Fig.\ \ref{state}(c) and the coupling 
$G(k,t)= \sqrt{2c_0\kappa/\pi}\sin(kL)\exp[i(\omega_0-\omega_k)t]$
is an \textit{odd} function centered on $\omega_k= \omega_g$. Thus, while $L$ satisfies the 
commensurability condition, \textit{i.e.,} $2L/c_0\! =\!2\pi n/\omega_g$ with 
$n\in \mathbb{N}$, 
the EC is in effect decoupled from the MC photon and the QD for the stationary state. 
As this occurs, the QD and the MC photon experience a cavity-assisted 
interaction at the rate of $\omega_g$, and both QD and MC-photon state 
are stationary when their complex amplitudes are $\pi$ out of phase. 
The only stationary state corresponding to the interaction Hamiltonian is antisymmetric in the 
amplitudes $c_e$ and $c_c$, and similar to the 
bound state in Ref.\ \cite{fang2015waveguide, calajo2018, TommasoPhysRevA.90.012113} or the subradiant/dark state in 
the recent studies of Refs.\ 
\cite{Zoller_1367-2630-14-6-063014, ZollerChiralPhysRevA.94.033829}. 

\section{Stability analysis}

In the previous section we identified a stationary state in the one-excitation subspace.
In this section, we ascertain the stationary state's stability, which depends on three important parameters. The first parameter is $n$,  related to $\omega_g$ and to the EC round trip time $\tau$ through the commensurability condition $\tau = n \tau_g$. The second one is the ratio $R = \kappa /4g$, between the cavity damping rate, $\kappa$ and the coupling strength $g$.  A third one is the feedback phase from the reflected photon after a round trip in the external cavity, $\Delta\phi = ( 4 \pi L / \Lambda_0 \mod{2\pi}) - \pi$. The reflected photon is assumed to 
undergo $\pi$ phase change from an ideal mirror ($r=-1$ in this study). We note that the exact value of $\omega_0$ plays a role in the system behavior only through this feedback phase, as $\omega_0 \tau = \Delta \phi + \pi \mathrm{\mod} 2\pi.$
 
To begin, we constructed a Markovian model to numerically simulate the 
evolution based on Eqs.\ (\ref{3})-(\ref{5}), and the 
stationarity of $\bar{\Psi}$ is
demonstrated in Fig.\ \ref{state_com}. 
For example, 
for $c_e(0)\!=\!1$, $c_c(0)\!=\!0$, and $c_k(0)\!=\!0$ for all $k$, 
the MC-photon population exhibits 
nonperiodic oscillations on the timescale of vacuum-field Rabi oscillations, 
shown in Ref. \cite{BiexcitonPhysRevLett.113.027401}. 

\begin{figure}[t]
    \centering
    \includegraphics[width=0.49\textwidth]{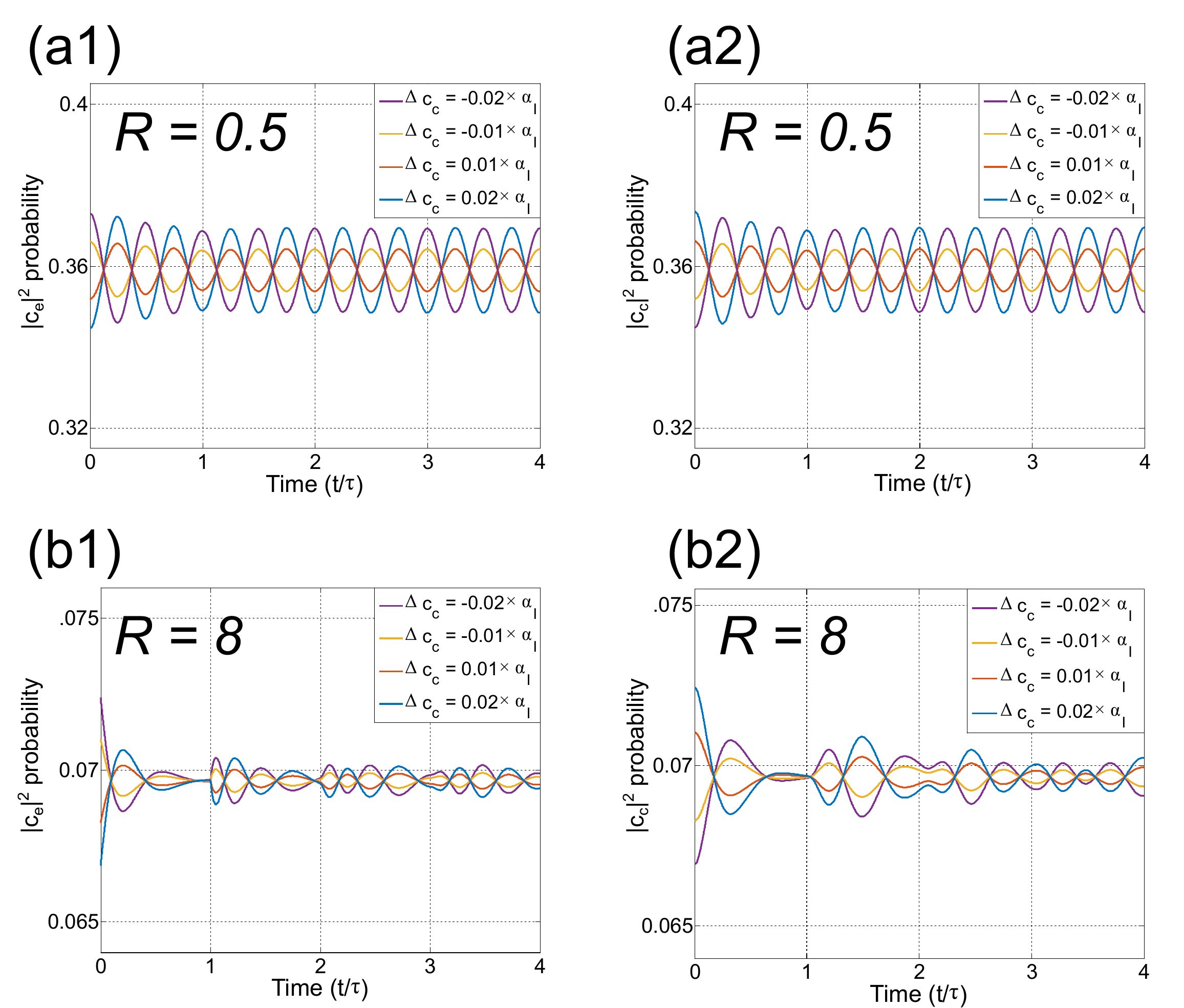}  %
    \vspace{-10pt}
    \caption{The evolution $\abs{c_e(t)}^2$ and $\abs{c_c(t)}^2$ of 
    perturbed stationary states with various values of $R$. 
    Note that the various probabilities remain near the initial values, indicating likely stability.} 
   \label{case1}
\end{figure}

Next, we study the dynamics in the vicinity of the stationary state to ascertain its stability.
We expect that for a stable state, the probabilities will remain near the initial values. 
Here, we perturb the stationary state and track the dynamics.  
(This numerical approach is infeasible \textit{rigorously}
to determine the stability of the candidate stationary 
state due to the existence of numerous degrees of freedom that 
would have to be perturbed individually; indeed, this is in effect what the method below of 
computing the eigenvalues of the Jacobian does for us.)
We add small values 
$\delta_{c,i}$ to  the initial conditions for $c_e$ and $c_c$ with respect to the stationary state 
$\bar{\Psi}(t)$ in Eqs.\ (6)-(8), \textit{viz.} 
$\delta c_c\! =\! [\pm~0.01, \pm~0.02]  \alpha$ 
while the amplitude of QD state $c_e(0)$ is perturbed such that 
$\abs{\bar{c}_e+\delta c_e}^2+\abs{\bar{c}_c+\delta c_c}^2 = 
\abs{\bar{c}_e}^2 + \abs{\bar{c}_c}^2$.
We shall see that the nature of the stable state depends crucially on the ratio 
$R \!=\! \kappa/(4g)$ between the MC damping rate $\kappa$ and the coupling strength $g$, 
the inversion of the conventional coupling strength parameter between a two-level system 
and a cavity, where $R<1$ ($R>1$) indicates the strong- (weak-) coupling 
regime \cite{strong_cavity_TLSPhysRevB.60.13276}.
Small $R$ thus means the vacuum-field Rabi frequency is much larger 
than the MC photon leakage rate;
large $R$ indicates a relatively high MC photon leakage rate
compared with the vacuum-field Rabi frequency. 
The effect on the stability is illustrated by plotting the probability 
for $R\!=\!0.5$ and $R\! =\! 8$. The probabilities of the excited QD 
state, $\abs{c_e(t)}^2$ are plotted in 
Fig.\ \ref{case1}(a1) and (b1) and MC photon state $|c_c(t)|^2$ in  
Fig.\ \ref{case1}(a2) and (b2) for $R \!=\!0.5,$ and $8$, respectively. Here, the feedback phase is $\Delta \phi = 0$ and $n=1$ in Fig.\ \ref{state_com} and all cases in Fig.\ \ref{case1}. 
 
In Fig.\ \ref{case1}(a1) and (a2), for small $R$ (strong coupling), 
the variation of the probability 
from its mean value exhibits oscillations at roughly the vacuum-field Rabi frequency 
$2\omega_g$ with some initial damping from 
$t \!=\! 0$ to $\tau$ at rate $2\kappa$. 
For $t\!\gtrsim \!\tau$, the damping is arrested
and oscillations at the vacuum-field Rabi frequency persist. 
By comparison, for $R\!=\!8$ [Fig.\ \ref{case1}(b1) and (b2)] when 
vacuum-field Rabi oscillations do not have a chance to
occur before photon leakage from the MC, the dynamics of both probabilities  are 
not evidently periodic,
indicating  the participation of multiple frequencies.  We shall see this is due to the
participation of many coupled EC-like modes.

In order to explore the stability of the stationary state in a rigorous fashion, we
consider an analysis of the eigenvalues of the Jacobian matrix \cite{krauskopf2000bifurcation}
to confirm the foregoing numerical simulations. 
The Jacobian is defined as the matrix of all first-order partial derivatives 
with respect to the each variable (in this case, the perturbed state amplitudes) 
evaluated for the stationary state  \cite{eckhaus1965studies}.  In other words,
the Jacobian probes the change in the stationary state with respect to an
arbitrary infinitesimal perturbation. Its eigenvalues, therefore, indicate whether or not
the state is stable, with a tendency to oscillate around the stationary state for 
eigenvalues being strictly \textit{imaginary}, stable and evolve toward a 
stationary state for \textit{real parts} of all eigenvalues being negative, 
or unstable and to evolve away from the stationary state
(all \textit{real parts} of any complex eigenvalues being positive). In addition, saddle points 
are also possible where the stationary state is stable against perturbations in 
certain directions in state space, but not in others.

More specifically, the Jacobian matrix $\boldsymbol{J_\delta}$ 
(see supplementary materials for definition)
satisfies the equation
\begin{flalign}
\frac{\partial }{\partial t} \delta c_i(t) = \boldsymbol{J_\delta}  \delta c_i(t)  . \label{10} \tag{10}
\end{flalign}
We begin by studying the Jacobian matrix constructed 
with linearized EOMs of the perturbed state. 
The perturbed states along the $i$ direction in state space from the 
stationary state 
will evolve according to $\delta c_i(t) = \delta c_i(0) e^{\lambda_i t}$ 
where $\lambda_i$ is the eigenvalue along the $i$ direction and $\delta c_i(t)$ is the small perturbation along the $i$ vector from the 
stationary state, $\bar{c}_{\alpha}$ at $t=0$. 
Given the eigenvalues, possible cases for the steady state are are 
attractors, repellors, or saddle points corresponding to all negative, 
all positive, or some positive and some negative eigenvalue, respectively. 
The dynamics can be analyzed by the nature of the equilibrium state, 
in this case, the stationary state, $\bar{\Psi}(t)$ in Eqs.\ (7)-(9). 

The interaction of the perturbed state amplitudes can be derived using 
linearized EOMs near the stationary state, and the Jacobian matrix of 
rank $(2+N)$ is derived following the small-signal model, 
where $N$ is the number of EC photon states, $c_k$, used in the simulation. 
Note that Jacobian, $\boldsymbol{J_\delta}$, is derived from linearization at the 
stationary state and is found to be a rank-$N+2$ skew Hermitian matrix,
while the interaction Hamiltonian is a rank-2 Hermitian matrix in the one-excitation subspace.

We proceed to analyze the stability of the stationary state by considering the 
eigenvalues, $\lambda_i$s, of $\boldsymbol{J_\delta}$. 
We focus on the dependence of the eigenvalues of $\boldsymbol{J_\delta}$ on 
the ratio $R \!=\! 4\kappa/g$
between the MC photon damping rate and the QD-MC coupling strength.
In addition, we carry out this analysis
adjusting two parameters, \textit{viz.}
the optical feedback phase $\Delta\phi$ and the EC round-trip time $\tau$. 
We first consider the effect of $\Delta\phi$, 
under conditions of constructive, 
destructive, and partial interference, and next 
consider the effect of varying $\tau$, restricting its value to
integer multiples of $\tau_g= \!2\pi/\omega_g$, 
\textit{i.e.,} $\tau = 2L/c_0\! =\!n\tau_g $
where $n\in \mathbb{\mathbb{N}}$.

In the above-mentioned scenarios, we find all the eigenvalues are purely 
imaginary ($i\lambda_i \in \mathbb{R}$). Imaginary eigenvalues indicate oscillatory dynamics upon perturbation about the stationary state, 
\textit{i.e.,} this indicates stability. Notably, since the perturbation 
of the stationary state evolves following 
$\delta c_c(t) = \delta c_c(0) \exp(\lambda_c t)$, we expect to observe 
oscillation in the probabilities both for the QD excited state and for the MC photon
$\abs{c_c(t)}^2 = \alpha^2 + 2 \alpha \delta c_c(0) \cos\omega_{\rm osc}t + 
\abs{\delta c_c(0)}^2$, where 
$\omega_{\rm osc}(\lambda)=\omega_0+\omega_g - i \lambda_c$. 
The eigenvalues of the Jacobian are solutions of the determinental equation
$|\boldsymbol{J_\delta}- \lambda  \mathrm{I_{(2+N_k)}}|=0$. One finds
\begin{flalign}
(\omega_{\rm osc}-\omega_g)^2 - \kappa (\omega_{\rm osc}-\omega_g)\sin(\omega_{\rm osc} \tau - \Delta \phi ) - \omega_g^2 = 0	. \label{11} \tag{11}
\end{flalign}
\noindent Note the relationship $\omega_0 \tau = \Delta \phi + \pi \mathrm{\mod} 2\pi$ is applied in the transformation,  accounting for the optical phase and the round-trip delay time. Where detailed derivation is presented in the Appendix C section. 

Here, we focus on finding the frequency $\omega_{\rm osc}(\lambda_c)$ as a function of 
the parameter $R = \kappa/(4g)$,
the feedback phase $\Delta\phi$ 
(obtained by varying $L$ on the lengthscale of the photon wavelength 
$\Lambda_0=2\pi c_0/\omega_0$) of the reflected photon, and and the time delay $\tau$. 
Given the wide range of $R$ investigated, 
the following results are plotted using a horizontal scale in $\log_2R$.

\begin{figure}[h]
    \centering
    \includegraphics[width=0.48\textwidth]{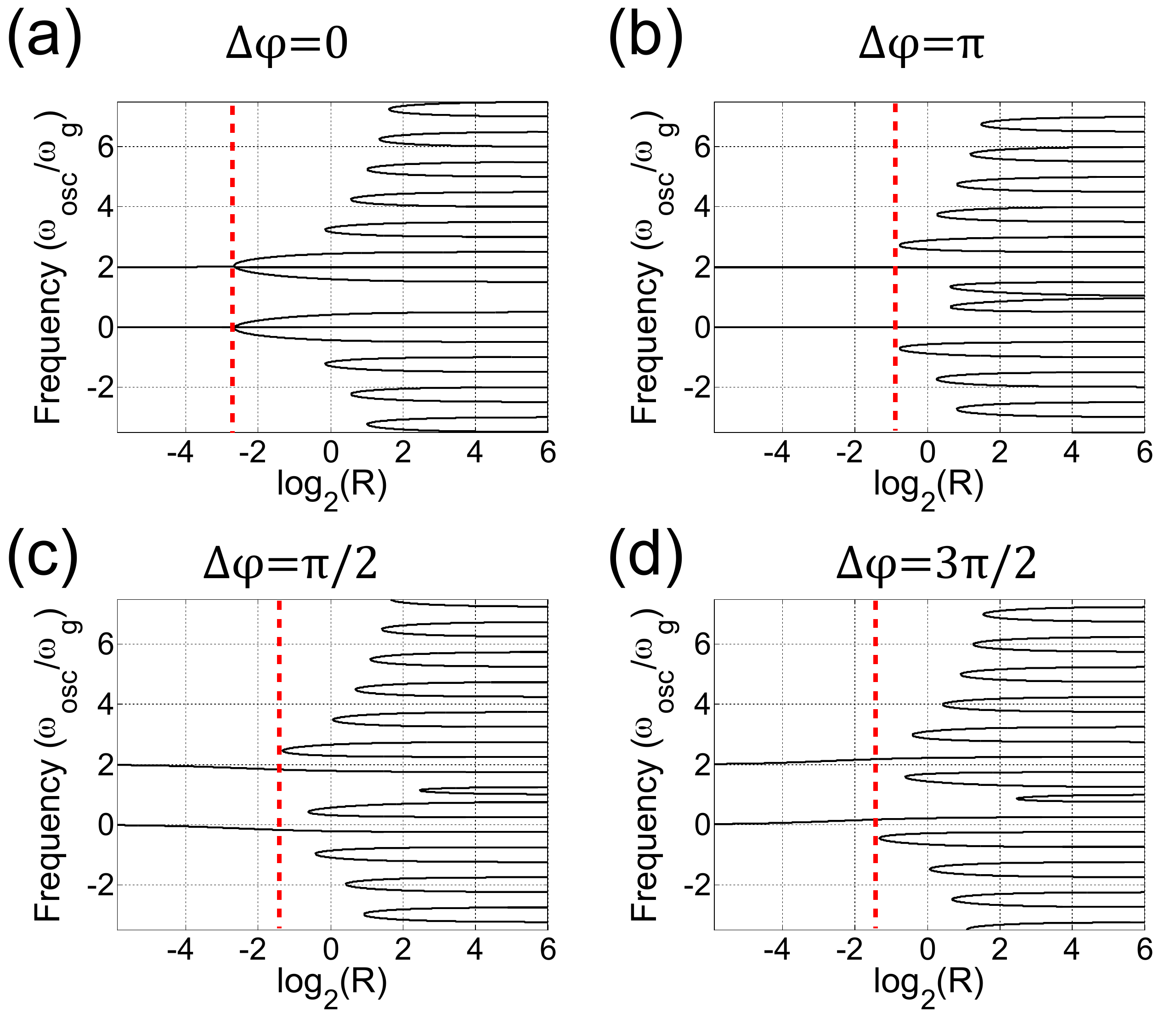}  %
    \vspace{-5pt}
    \caption{Frequencies of oscillation $\omega_{\rm osc}$ about the stationary state of the 
    probability $\abs{c_c(t)}^2$ as a function of  $\log_{2}R$ for various $\Delta\phi$, while n=1. The frequency is obtained from 
    $\omega_{\rm osc}=\omega_g+\omega_0- i \lambda$, where $\lambda$ is strictly imaginary. } 
       \vspace{-10pt}
   \label{Fre_BD_ana2}
\end{figure}

\begin{figure}[h]
    \centering
    \includegraphics[width=0.5\textwidth]{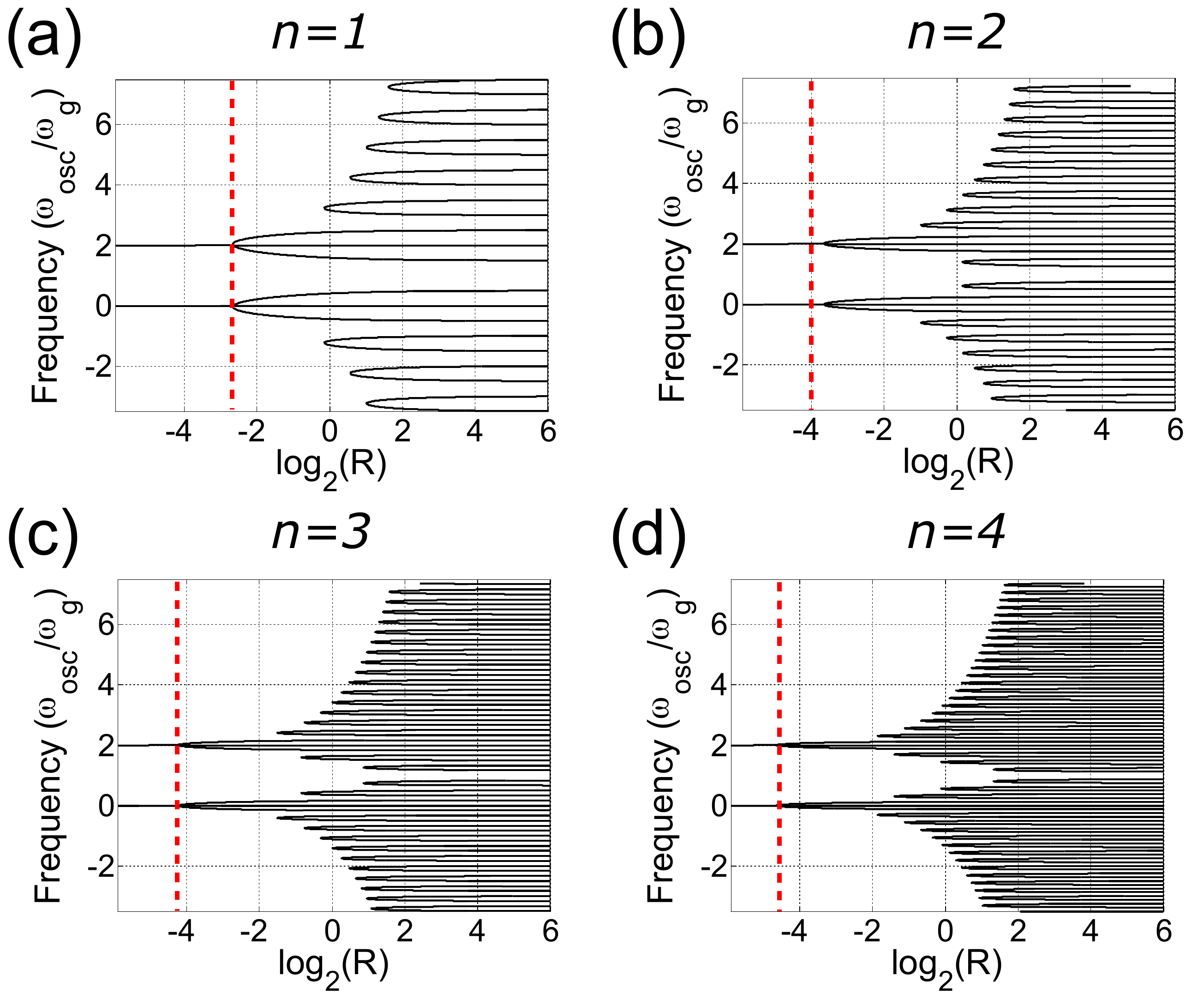}  %
    \vspace{-10pt}
    \caption{Frequencies of oscillation $\omega_{\rm osc}$ about the stationary state 
    of the probability $\abs{c_c(t)}^2$ as a function of $\log_{2}R$ 
     for various $\tau_g$ in the case where the feedback phase, $\Delta\phi = 0$. 
     The frequency is obtained from 
    $\omega_{\rm osc}=\omega_g+\omega_0- i \lambda$.    } 
       \vspace{-10pt}
   \label{Fre_BD_nL2}
\end{figure}

\noindent \textbf{Constructive interference}: $\Delta\phi\! =\! 0$. The 
EC round-trip distance is a half-integer multiple of the photon wavelength, 
$2L \!=\! (m+1/2) \Lambda_0$.  
For $R$ less than a critical value $\bar{R}(n, \Delta \phi)$ (depending on $\Delta\phi$ and $\tau$),
we find only one non-zero imaginary eigenvalue and it 
corresponds to oscillatory dynamics about the stationary state (\textit{i.e.,} stability) 
at frequency  $2\omega_g$. 
In this region, starting with perturbed initial conditions results in limit-cycle dynamics.
An arbitrary perturbation leads to oscillatory dynamics at the Rabi frequency.
For values of $R>\bar{R}(n, \Delta \phi)$, we find multiple emerging imaginary 
eigenvalues as we increase $R$. 
In this region, for large $L$ we have in effect a quasi-continuum of EC modes, and
an arbitrary perturbation leads to non-oscillatory dynamics involving the many frequencies 
associated with the various eigenvalues of the Jacobian.  This is a region of asymptotic stability.
Recall that $n$ is the ratio of the EC round-trip time $\tau$ and 
the Rabi period $\tau_g=2\pi/\omega_g$.
$\bar{R}(n, \Delta \phi)$ is indicated by the vertical dotted 
red line in Figs.\ \ref{Fre_BD_ana2} and \ref{Fre_BD_nL2}.  
In the case of constructive interference, 
for $R=\bar{R}(1, 0)\!=\! 0.156$ 
($\log_2R=-2.68$) as shown in Fig.\ 5(a),
we find six solutions, indicating a possible bifurcating point in the strong-coupling regime.  
We also find that additional oscillatory modes emerge pairwise and 
their eigenfrequencies are smooth functions of $R$ satisfying Eq. (11). 
In addition, the frequency separation between the additional eigenvalues 
in the weak-coupling regime \textit{e.g.} $R \!= \!64$
($\log_2R=6$), tends toward $\delta f_{\rm osc} = 1/(2\tau)$, the EC free spectral range. 
The expected oscillation frequencies for $n=1$ as a function of $R$ are shown in 
Fig.\ \ref{Fre_BD_ana2}(a).

\noindent \textbf{Destructive interference}: $\Delta\phi \!= \!\pi$. The 
EC round-trip distance is an integer multiple of the photon wavelength, 
$2L\! =\! m \Lambda_0$. The reflected photon interferes 
destructively with the emitted photon at the MC.
We plot the oscillation frequencies as a function of $\log_{2}R$ in 
Fig.\ \ref{Fre_BD_ana2}(b)
for $n=1$. Additional eigenvalues appear for $R$ above a 
different critical value $\bar{R}(1, \pi)\! =\! 0.62$
($\log_2\bar{R}(1,\pi)=-0.69$, lower than that for the case of constructive interference)
as shown in Fig.\ 5(b). 
For $R$ increasing above $0.62$, two pairs of oscillating modes 
emerge (in addition to the existing solutions, $\omega_{\rm osc} = 0$ and $2\omega_g$).
The higher-frequency pair of eigenvalues emerges at 
approximately $-i(\omega_0  + \omega_g) + i 2.75\omega_g$, 
and further splits into values separated by the EC free spectral range for higher 
values of $R$ (weaker coupling strength). 
Note that in both cases of constructive  and destructive interference, 
the additional solutions for the frequency, 
$\omega_{\rm osc}$, appear symmetrically with respect to 
the $\omega_{\rm osc}=\omega_g$ axis from Eq.\ (11).

\noindent \textbf{Partial interference}:  $\Delta\phi \!\neq \!0\! \neq \!\pi$.
In this case, the reflected photon partially interferes with the emitted photon. 
The oscillation frequency of the MC photon state probability, 
$\omega_{\rm osc}$ depends also on the 
phase difference of the reflected photon. We plot the MC oscillation frequencies, 
$\omega_{\rm osc}$  for $\Delta\phi \!=\! \pi/2$ and $3\pi/2$ in 
Fig.\ \ref{Fre_BD_ana2}(c) and (d), respectively. 

An analogy exists between the behavior of the perturbed dynamics about 
the stationary state in the quantum system under consideration and 
the nonlinear dynamics of the ECSL. Under weak feedback, the LK equations predict
the ECSL dynamics are strongly influenced by the optical feedback phase. 
Namely, it is indicated in Ref. \cite{tkach1985linewidth} that 
\textit{the laser line does not split under the influence 
of out-of-phase feedback, but does alternately narrow and 
broaden depending on the phase of the feedback.} Since weak feedback in the 
ECSL is analogous to $R>\bar{R}$ in the quantum system, 
we comment here on the optical phase of the feedback field. 
Specifically, the feedback phase of \textit{a single photon} has a similar 
impact on the dynamics in the weak coupling regime for the MC/QD/EC system. 
These effective changes appear near $R$ slightly above different critical values of 
$\bar{R}$ under the various interference conditions shown in Fig.\ 5(a)-(d). 

We next explore the stability dependence on $R$ for various $L$. 
We find the stationary state only exists when commensurability condition is satisfied, \textit{i.e.,} the time delay, $\tau=n\;\tau_g$, is an integer multiple of $\tau_g$ where  $\tau_g\!=\! 2\pi/\omega_g$, and $n\in\mathbb{N}$. 
We chose $\Delta\phi \!=\! 0$ in all cases in Fig. \ref{Fre_BD_nL2}.

In the case of coherent quantum feedback from a single photon, 
one can show that the \textit{product} of the time delay $\tau$ 
and the dimensionless ratio $\bar{R}(n, 0)$ is constant $1/(2\pi)$ for $n\in\mathbb{N}$ from Eq. (11). We find the critical value $\bar{R}$ of $R$, beyond which more than two (imaginary) eigenvalues of the 
Jacobian matrix appear decreases as $L$ increases shown by various 
vertical dotted red lines in the panels of Fig. \ref{Fre_BD_nL2}. 
That is, $\bar{R}(1, 0) = 0.16$
[$\log_2\bar{R}(1, 0) =-2.74$], 
$\bar{R}(2, 0) = 0.08$
[$\log_2\bar{R}(2, 0) =-3.64$],
$\bar{R}(3, 0) = 0.053$
[$\log_2\bar{R}(3, 0) =-4.24$],
and $\bar{R}(4, 0) = 0.039$
[$\log_2\bar{R}(4, 0) =-4.68$].

In the LK model for an ECSL, this product is also proportional to the dimensionless 
parameter $C = \gamma \tau$ characterizing feedback strength
defined in Ref.\ \cite{LENSTRA1984255}, where $\gamma$ is the \textit{feedback parameter}.
Specifically, in the theoretical study of weak optical feedback (classical ECSL), 
$C <1$ there is always \textit{one} stable solution; 
for $C>1$ there may exist more stable solutions 
corresponds to single-frequency operation \cite{LENSTRA1984255}.
The frequency difference between the multiple solution 
of $\omega_{\rm osc}$ scales linearly with $n$ as 
one expected from the LK model. As we pointed out in the previous paragraph, the critical value, $\bar{R}$ in the single-photon limit corresponds up to a constant multiplicative factor to the feedback parameter, $\gamma$ in the classical regime. That is, $\bar{R}$ satisfies the relationship  $ n \bar{R} = 1/(2\pi) $ ($ \tau\bar{R} = \tau_g /(2\pi)$), represents the feedback parameter in the single photon regime.

\section{Discussion and conclusion}

The implementation of coherence feedback with few-excitation states incorporating 
cQED systems can be realized in various ways. Albert, \textit{et al.,} have 
realized coherent optical feedback with microlaser, where chaotic behavior is 
observed with self-feedback for a few few photons $\sim 100$, and studied 
the second-order autocorrelation function $g^{(2)}(\tau)$ \cite{QDchaosalbert2011}. 
Note that in these experiments, more realistic parameters, 
including the QD dephasing rate, 
the QD decay rate, the transmission and reflection coefficients of the 
EC mirror and the MC-EC coupling are needed. 
However, these parameters describe the loss channels and contribute 
to decay of the probabilities, we expect these parameters will weakly influence the 
oscillation frequencies provided the large $R$ limit can be reached.

 A stability analysis of the type implemented here can also be applied to the 
dark state existing in a quantum dimer system with a coupled 
cavity \cite{Zoller_1367-2630-14-6-063014, ZollerChiralPhysRevA.94.033829} or 
in certain spin systems \cite{dimer_PhysRevA.93.062104, zoller_PhysRevA.91.042116}. 
In the case with spin system, the coupling element between two interacting spins 
placed at distance $L$ apart is replaced by 
$G_{sd}(k',t)= G_0\cos(k'L)e^{i(\omega_0-\omega_k)t}$ \cite{dimer_PhysRevA.93.062104}. 
By making the transformation $k' = k-\pi/2$, we obtain the relationship between the 
eigenvalues in two cases, \textit{i.e.}, the eigenvalues of $\lambda_{sd} = \lambda-\pi/2$. 
Following the derivation of this work, we can solve the emergent behavior 
of the additional solutions of 
$\omega_{osc, sd}(R,\Delta\phi,n) = \omega_{\rm osc}(R,\Delta\phi - \pi/2,n)$ 
near the stationary (dark) state. 
Thus, the dynamical behaviors of the interacting quantum dimer coupled 
through (single) photon bath should behave similarly to coherent quantum feedback. 
In this case, the frequencies appearing in Fig.\ 5 (a) and (b) must be interchanged as well as those appearing in Fig. 5 (c) and (d).

In conclusion, we consider a system composed of a QD in a MC 
coupled to a EC and find a stationary state where the state 
initialized in the MC and QD degrees of freedom is stable against decay into
EC photons.
Specifically, we give the analytical expression for the stationary solution 
for such a system in the one-excitation subspace. 
We find this state is stable 
by performing stability analysis on the Jacobian of state amplitudes. 
The periodic solutions perturbed about the stationary state, obtained 
from the eigenvalues of Jacobian, 
indicate additional solutions arise above a critical value $\bar{R}$ or $R$ as 
experimental parameters such as the cavity damping rate, the cavity coupling strength, 
the feedback phase, and the external cavity length are varied.
We found a strong similarity to the LK model in this behavior in terms of its dependence
on $\Delta\phi$ and $\tau$. In addition, numerical simulation verifies these results, 
showing interesting dynamics appear in the vicinity of the stationary states. 
Our stability analysis may serve as a bridge between classical and quantum 
models for nanophotonic structures subject to optical feedback.

\section{Acknowledgment}
We thank Dr. Yao-Lung L. Fang, and Prof. Pascale Senellart for useful discussions. The authors gratefully acknowledge the financial support of the 2015 Technologies Incubation scholarship for Dr. C. Y. Chang from Ministry of Education, Taiwan.

\end{document}